\documentclass{article}

\usepackage{arxiv}

\usepackage[utf8]{inputenc} % allow utf-8 input
\usepackage[T1]{fontenc}    % use 8-bit T1 fonts
\usepackage{hyperref}       % hyperlinks
\usepackage{url}            % simple URL typesetting
\usepackage{booktabs}       % professional-quality tables
\usepackage{amsfonts}       % blackboard math symbols
\usepackage{nicefrac}       % compact symbols for 1/2, etc.
\usepackage{microtype}      % microtypography
\usepackage{lipsum}
\usepackage{graphicx}
\graphicspath{ {./img/} }

\usepackage{amsmath,amssymb,amsfonts}
\usepackage{tabularx}
\usepackage{multirow}
\usepackage{subfigure}

\title{ScalingNet: extracting features from raw EEG data for emotion recognition}

\author{
 HU Jingzhao \\
  Department of Information Science and Technology\\
  Northwest University\\
  Xi’an 710127, China \\
 \And
 WANG Chen \\
  Department of Information Science and Technology\\
  Northwest University\\
  Xi’an 710127, China \\
 \And
 JIA Qiaomei \\
  Department of Information Science and Technology\\
  Northwest University\\
  Xi’an 710127, China \\
 \And
 BU Qirong \\
  Department of Information Science and Technology\\
  Northwest University\\
  Xi’an 710127, China \\
 \And
 FENG Jun$^*$ \\
  Department of Information Science and Technology\\
  Northwest University\\
  Xi’an 710127, China \\
  \texttt{fengjun@nwu.edu.cn} \\
}

\begin{document}
\maketitle
\begin{abstract}
Convolutional Neural Networks(CNNs) has achieved remarkable performance breakthrough in a variety of tasks. Recently, CNNs based methods that are fed with hand-extracted EEG features gradually produce a powerful performance on the EEG data based emotion recognition task. In this paper, we propose a novel convolutional layer allowing to adaptively extract effective data-driven spectrogram-like features from raw EEG signals, which we reference as scaling layer. Further, it leverages convolutional kernels scaled from one data-driven pattern to exposed a frequency-like dimension to address the shortcomings of prior methods requiring hand-extracted features or their approximations. The proposed neural network architecture based on the scaling layer, references as ScalingNet, has achieved the state-of-the-art result across the established DEAP benchmark dataset. 
\end{abstract}

% keywords can be removed

%% main text
	\section{Introduction}
		%导入、相关、引出、贡献、总结
		% 注意，每个地方几句话，不要写太多
		% todo 导入领域和问题、按技术流派简要汇总、【通过已有方法的问题引出自己的工作（也可以分散在第二部分中）、简述动机、原理、阐明贡献、汇报结果】（方括号内胡景钊写）
		Emotion recognition plays a very important role in human-computer interaction\cite{li2017human}. Through recognizing human emotions more accurately and quickly, we can promote a smarter life\cite{liu2016multimodal}. Generally, expressive modalities are used to judge human being's emotions, such as facial expressions, audio-visual expressions, and body language, etc.\cite{adolphs1994impaired}. In recent years, more and more studies that recognize human emotions have used physiological electrical signals\cite{issa2020emotion}\cite{gao2020gpso}, such as electrocardiogram (ECG), electromyography (EMG) and electroencephalography (EEG). Among them, EEG signals can better reflect real human emotions because it is not affected by subjective factor\cite{mokatren2019improved}. In this work, we use EEG signals to recognize human emotions.

        It has been proved that there are intimate correlations between human emotions and their different brain states\cite{gao2020channel}\cite{yang2019multi}. With the progress in EEG hardware equipment, it is more convenient to collect EEG signals with a high sampling rate nowadays\cite{fabiano2019emotion}. Meanwhile, the processing and analysis methods of EEG signals are being explored and researched constantly\cite{li2019eeg}. In EEG based emotion recognization, researchers mainly focus on three technical sects. The first and most widespread methods are based on feature engineering and machine learning algorithms to recognize human emotions \cite{zheng2019emotionmeter:}, which requires hand-extracted emotion-related features from EEG signals, such as Power Spectral Density (PSD), Differential Entropy (DE), etc. With the proposal of deep learning, some methods tend to combine feature engineering and deep neural networks, which replace classifiers from machine learning algorithms to deep neural networks, such as CNNs\cite{xing2019sae+lstm:}. Furthermore, some researchers consider extracting data-driven features from EEG signals, which employ parameterizable data representation methods or neural networks\cite{chen2019a} as a feature extractor. While the feature extraction methods mentioned above achieved remarkable performance of EEG based emotion recognition, there is still potential for improvement. Hand-extracted features are mostly tasks related, and mostly require strong hypotheses and mathematically driven theoretical supports. Considering the reality, we may say that the hand-extracting of features is not easy works and potentially not robust.
        
		% todo:接着可以从现有的SVM、LSTM和CNN等引入到ScalingNet方法，简述咱们的动机原理等。
		Inspired by the shortcomings of hand-extracted feature based methods, we introduce an end-to-end artificial neural network method mainly constructed by our well-designed data-driven signal feature extracting layer, which we reference as ScalingNet, allowing to robustly performs raw EEG data based emotion recognition without requiring any hand-extracted features. The idea of the layer, which we reference as scaling layer, is to dynamically generate a series of convolution kernels scaled from one data-driven pattern to produce a robust data-driven spectrogram-like feature map from raw EEG signals for downstream tasks. The introduced architecture has several interesting properties:(1)It automatically extracts robust feature maps from raw EEG signals without any hand-interaction. (2)It handles any length of EEG signal without requiring data alignment. (3)It is fully convolutional. (4)It is compatible with the existing neural networks, providing robust feature extraction for different downstream tasks. We validate the proposed approach on the challenging DEAP benchmark dataset, achieving the state-of-the-art result that highlights the potential of models for data-driven feature extraction from raw EEG signals.
		
	\section{Related Work}
		%按逻辑总结相关工作、【存在问题并引出自己工作的必要性】（方括号内为可选）
		% todo 可大体按照引言的第二部分逐技术流派分段展开
		%In recent years, EEG signals are growly used in emotion recognition because of its objectivity, which can obtain real information about human emotions. Zheng et al.\cite{zheng2016identifying} extracted the time domain and frequency domain features of EEG, such as differential entropy(DE), Power Spectral Density(PSD), etc., and used Support Vector Machine (SVM) for emotion classification. Liu et al.\cite{liu2016emotion} extracted time domain, frequency domain and time-frequency domain features, such as Hjorth, PSD, Discrete Cosine Transform(DCT), etc., use k-Nearest Neighbor(KNN) and Random Forest(RF) as classifiers for classification. Li et al.\cite{li2016emotion} proposed to perform Continuous Wavelet Transform(CWT) on the EEG signal of each channel, and then convert it to scalograms, and input the construction frame into CNN and Long Short-Term Memory (LSTM) for emotion recognition. For the classification effect, Kim et al.\cite{kim2018deep} proposed to extract brain asymmetry features and heart rate features, respectively, and ConvLSTM(Combination of CNN and LSTM) was used for classification. Wang et al.\cite{wang2018emotionet:} proposed an EmotionNet network for EEG-based emotion classification. It takes EEG as input and uses 3-D convolution to extract spatial and temporal features for emotion recognition.
		
		In EEG based emotion recognition, machine learning based methods fed with hand-extracted EEG features are possibly the most widely used framework. With the development of deep learning, researchers gradually tend to replace machine learning methods with deep neural networks, especially CNNs\cite{roy2019deep}. The hand-extracted EEG features are mainly time domain, frequency domain, time-frequency domain and spatial signal features. The classification methods mainly include random forest, SVM, CNNs, LSTM, etc. Zheng et al.\cite{zheng2016identifying} extracted the time domain and frequency domain features of EEG, such as differential entropy(DE), Power Spectral Density(PSD), etc., and used Support Vector Machine (SVM) for emotion classification. Liu et al.\cite{liu2016emotion} extracted time domain, frequency domain and time-frequency domain features, such as Hjorth, PSD, Discrete Cosine Transform(DCT), etc., use k-Nearest Neighbor(KNN) and Random Forest(RF) as classifiers for classification. Li et al.\cite{li2016emotion} proposed to perform Continuous Wavelet Transform(CWT) on the EEG signal of each channel, and then convert it to scalograms, then input the construction frame into CNNs and Long Short-Term Memory (LSTM) for emotion recognition. Kim et al.\cite{kim2018deep} proposed to extract brain asymmetry features and heart rate features, respectively, and ConvLSTM(Combination of CNN and LSTM) was used for classification. 
		
		Inspired by the powerful feature transform ability of deep neural networks, some researchers commit to design an end-to-end framework for EEG based emotion recognition. Wang et al.\cite{wang2018emotionet:} proposed an EmotionNet network for EEG-based emotion classification. It can take EEG as input and uses 3-D convolution to extract spatial and temporal features for emotion recognition. However, for general purpose network layers, it is hard to learn and extract robust features from signals. In the long run, this research field still has great potential for development. Consider, there is a need for a special neural network layer that specially design for robust feature extraction from raw EEG signals,  and a neat neural network architecture that can naturally inference on raw EEG signals.

	\section{Methodology}
		% 方法胡景钊来写
		% todo 一段话来总领本章
		In this section, we will firstly present the building block layer used to adaptively extract effective data-driven spectrogram-like features from raw EEG signals, which we reference as scaling layer. Then we will introduce a fully convolutional neural network constructed through basing the scaling layer, which we reference as ScalingNet since its core feature is the application of scaling layer.
		
		\subsection{scaling layer}
			% todo 一段话简述动机、原理、目的或意义
			The motivation is to dynamically generate a series of convolutional kernels by scaling one data-driven pattern to different periods in order to expose a frequency-like dimension from signals. This motivation brings the possibility of automatically adaptive extracting effective and robust data-driven spectrogram-like features for downstream tasks from raw EEG signals.
			
			% todo 从假设或问题开始描述方法，还要附图
			We consider a multi-kernel convolutional layer that takes a one-dimensional signal shaped like $(sampling\ points, 1)$ as input and a two-dimensional spectrogram-like feature map shaped like $(sampling\ points, scaling\ levels)$ as output with the following defined layer-wise propagation rule:
			
			\begin{equation}
				\label{equ:sl}
				H^{output}(l) = \delta(bias(l) + downSample(weight, l) \otimes H^{input})
			\end{equation}
			
			where $H^{input}$ is the input vector shaped like $(time\ steps, 1)$, i.e. the one-dimensional signal. $H^{output}$ is the matrix of activations shaped like $(time\ steps, scaling\ levels)$, i.e. the data-driven spectrogram-like feature map. $bias$ is the biases for multi-kernel generated by scaling a basic kernel. $\delta(\cdot)$ denotes an activation function; $weight$ is the basic kernel where others kernel scaled from. $l$ is a hyper-parameter that controls the scaling level. 
			
			$\otimes$ is a valid cross-correlation operator, normally defined as:
			
			\begin{equation}
				\label{equ:cc}
				(f \otimes g)[n] \triangleq \sum_{m=0}^{N-1}\overline{f[m]}g[(m+n)_{\mod N}]
			\end{equation}
			
			where $f$ is $downSample(weight, l)$, $g$ is $H^{input}$.
			
			$downSample$ is a pooling operator that downsamples the $weight$ by average filter with a window of size 2. To ensure that the length of downsampled $weight$ still is odd, the $downSample$ setup a padding of size 1 for the filter when the length of directly downsampled $weight$ potentially is even.
			
			Further, $bias(l)$ is the bias for the kernel generated at $l^{th}$ scaling level. $H^{output}(l)$ is the activation of $l^{th}$ scaling level. $downSample(weight, l)$ denotes the generated kernel scaled from $weight$ at $l^{th}$ level, which recursively filters the $weight$ $l$-times.
			
			Steply, assume we would extract features for signal $H^{input}$ at $l^{th}$ scaling level. We first generate the $l^{th}$ scaling level kernel scaled from $weight$ by $downSample(weight, l)$. Then, we perform the cross-correlation operator of the scaled kernel and $H^{input}$ by Equation \ref{equ:cc}. Then, we add the previous result and the $bias(l)$, and then feed to activation function $\delta(\cdot)$, i.e. Equation \ref{equ:sl}. 
			
			We repeat the above process expected total scaling level times with different setup of hyper-parameter $l$ on a range of 0 to maximum scaling level. Finally, we stack all extracted feature vectors into a 2D tensor to obtain the data-driven spectrogram-like feature map. In particular, in order to ensure the alignment of extracted feature vectors, the length of basic kernel $weight$ must be odd and the input signal $H^{input}$ must be padded with $(scaledKernelLength - 1)/2$. For the backpropagation, the trainable parameter are the basic kernel $weight$ and biases $bias$, which will be handled by autograd mechanism.
			
			\begin{figure}[htbp]
				\centering{\includegraphics[width=0.95\textwidth]{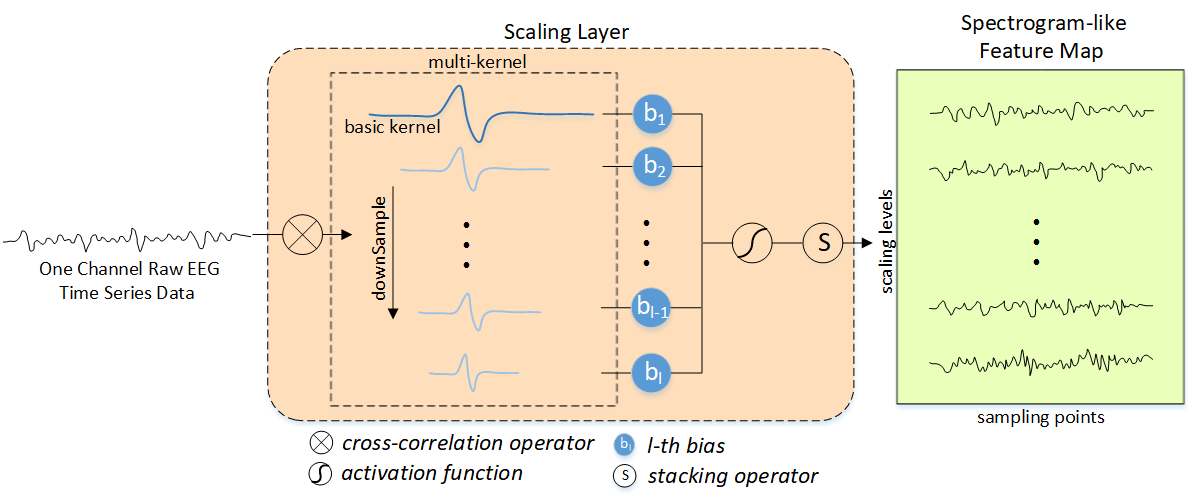}}
				\caption{The core principle of scaling layer. Scaling layer directly extracts data-driven spectrogram-like feature maps from raw EEG signals for downstream tasks. It extracts feature by multi-kernel generated from scaling a data-driven pattern.}
				\label{fig:sl}
			\end{figure}
			
			% todo 这里应该有一个sl示意图
			The core principle of scaling layer is illustrated by Figure \ref{fig:sl}.
			
			% todo 可选地： 介绍算法，1.引出并给出算法流程。2.对算法进行一定的解释。
			
		\subsection{ScalingNet}
			% todo 一段话，启下、承上、应用点
			In this subsection, we introduce a neural network architecture mainly constructed by a series of parallel scaling layers to perform raw EEG data based emotion recognition, which we reference as ScalingNet.
			
			\begin{figure}[htbp]
				\centering{\includegraphics[width=0.95\textwidth]{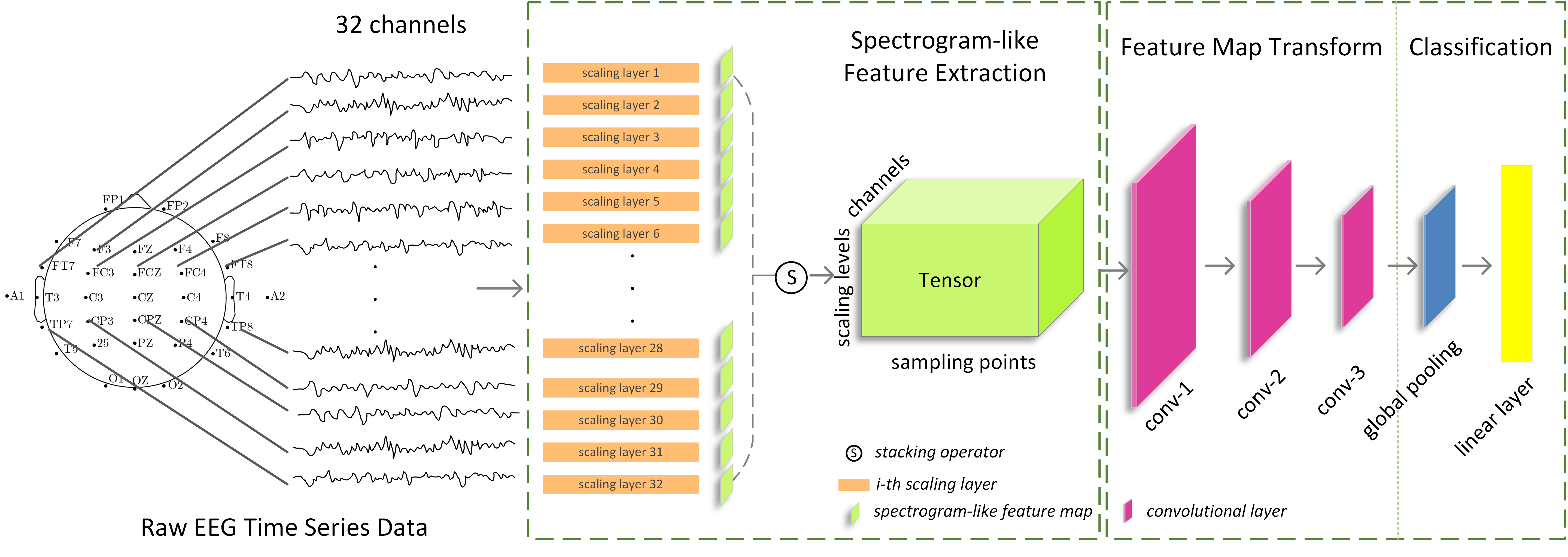}}
				\caption{The ScalingNet architecture. It's mainly constructed by a series of parallel scaling layers that are followed by neat convolutional and linear layers. With the help of data-driven spectrogram-like feature maps extract by scaling layers, it performs raw EEG data based emotion recognition without any hand-extracted features.}
				\label{fig:sn}
			\end{figure}
			
			The ScalingNet architecture is illustrated in Figure \ref{fig:sn}. Considering that the scaling layers that mainly used to construct the ScalingNet extract data-driven spectrogram-like feature maps for EEG channels separately, we especially illustrate the EEG channels by carefully stacking the data-driven spectrogram-like feature maps extracted by scaling layer from EEG signal of different channels into a 3D tensor.
			
			The EEG signals of different channels are first fed to scaling layers separately to extract data-driven spectrogram-like feature maps. Then, the feature maps extracted by scaling layers are stacked into a 3D tensor along the EEG channel dimension. Then the 3D tensor fed into several convolutional layers to perform feature map transform. Finally, the transformed features maps are fed into an average global pooling layer and a linear layer to perform emotion classification. Worthily, the ScalingNet architecture robustly performs raw EEG data based emotion recognition without requiring any hand-extracted features.
			% todo 附图并解释（把网络结构受清楚就行）
			
	\section{Experimental \& Results}
		%数据集介绍、数据预处理、依次编排实验介绍/设置/结论、按需看是否需要总结
		% todo 一段话总领本章，如何比较、什么数据集、结果如何、本章安排
		We evaluate the performance of the proposed ScalingNet architecture on EEG data based emotion recognition task using the established challenging DEAP dataset\cite{koelstra2012deap:} and compare it with strong benchmarks or previous state-of-the-art methods. In this section, we first introduce the DEAP dataset, then proceed to a detailed description of the experimental setup, and finally report the experimental results.
				
		\subsection{Datasets}
			% todo 详细描述数据集，什么数据，已经如何被处理了，数据集的含义与标签、训练集测试机如何划分、数据规模
			%? In this paper, we used 32 channels of EEG signals and the three dimensions of valence, arousal, and dominance for emotion recognition. We choosed 5 as the rating threshold\cite{8489331} and treated the task in this paper as three binary classification problems. We used a total of 1280 samples of 32 subjects for cross-subject emotion recognition, and would use cross-validation in the experiment to evaluate the performance of the method proposed in this paper.? The DEAP dataset consisted of 32 subjects which were downsampled to 128 Hz and band filtered from 4-45 Hz. Each subject watched 40 music videos, and each video would be rated 1-9 in term of valence, arousal, dominance and liking\cite{8512865}. The 63s signal was recorded for each video, removing the first 3s of baseline data. In this paper, we used 32 channels of EEG signals and the three dimensions of valence, arousal, and dominance for emotion recognition. We choosed 5 as the rating threshold\cite{8489331} and considered the task in this paper as three binary classification problems.
			
			The DEAP\cite{koelstra2011deap} is an established challenging benchmark dataset for EEG based emotion recognition. The dataset contains EEG and physiological signals collected from 32 subjects stimulated by watching music videos. After they watch each video, they self-evaluate their valence, arousal, dominance, and liking according to 1-9 immediately. Each subject is asked to watch 40 videos, and 63 seconds of signals are collected for each video. The signals are default downsampled to 128Hz and filtered with a 4.0Hz to 45.0Hz bandpass filter. In this paper, only EEG signals are used to classify the valence, arousal, and dominance by the rating threshold of 5, which closely follows the setting of \cite{8489331}. Specifically, 1280 EEG samples from 32 subjects are used for three binary classification tasks of cross-subject emotion recognition.			
			
		\subsection{Experimental setup}
			% 讲讲网络的超级参数设置，等
			% todo 网络结构的细节
			
			% todo 网络里的参数，与细节一览表：表的解释与表
			The five-fold cross-validation strategy is employed to objectively evaluate the raw EEG data based emotion recognition performance of the proposed ScalingNet architecture. We manually optimize the hyper-parameters of proposed ScalingNet architecture on the DEAP dataset, and the most related tuned hyper-parameters are reported in Table \ref{tab:hyperparameters}. where the "length of $weight$" is the size of basic kernel $weight$ of scaling layer in Equation \ref{equ:sl}. The "kernel size" is the size of convolutional kernels used in feature map transform convolutional layers of proposed ScalingNet architecture illustrated in Figure \ref{fig:sn}. The "number of filter" is the number of filters used in feature map transform convolutional layers of proposed ScalingNet architecture illustrated in Figure \ref{fig:sn}.
			
			\begin{table}[htbp]
				\centering
				\caption{The hyper-parameters of proposed ScalingNet architecture tuned on the DEAP dataset.}
				\label{tab:hyperparameters}
				\begin{tabularx}{0.95\textwidth}{c | >{\centering\arraybackslash}X}
					\toprule
					Hyper-parameters & Value \\
					\midrule
					batch size & 32 \\
					length of $weight$  & 33 \\
					kernel size & $3 \times 5$ \\
					number of filter & 16, 8, 6 \\
					activation function & relu \\
					loss & cross entropy \\
					optimizer & adam \\
					\bottomrule
				\end{tabularx}
			\end{table}
			
			% 这里写实验环境pyotrch、nni、2080ti
			All experiments in this paper were conducted using a Geforce RTX 2080 Ti. The machine learning framework used in this paper is PyTorch\cite{pytorch}.
		
		\subsection{Results}
			% todo 说说怎么评估、对照数据都是怎么得来的、以及最终的结果、并下结论、看情况是需要重申原理（还是放在讨论部分）
			
			The experimental results of the proposed ScalingNet architecture compared with previous state-of-the-art methods using the DEAP dataset and the same evaluation strategy are shown in Table\ref{tab:result}. Where evaluation criteria are the emotion recognition accuracies of arousal, valance, dominance in closely following previous studies. 	
			
%			\begin{sidewaystable}[htbp]
			\begin{table}[htbp]
				\centering
				\caption{Experiment results compared with previous state-of-the-art methods.}
				\label{tab:result}
				\begin{tabularx}{0.95\textwidth}{c | >{\centering\arraybackslash}X c >{\centering\arraybackslash}X >{\centering\arraybackslash}X >{\centering\arraybackslash}X}
					\toprule
					\multirow{2}{*}{Studies} & \multirow{2}{*}{Features} & \multirow{2}{*}{Classifiers} & \multicolumn{3}{c}{\textbf{Accuracy}} \\
					%\cline{4-6} 
					 &  &  & Arousal & Valance & Dominance \\
					\midrule
					Li et al. & DBN & SVM & 0.6420 & 0.5840 & 0.6580 \\
					CHEN et al. & - & H-ATT-BGRU & 0.6650 & 0.6790 & - \\
					Yang et al. & VAE & SVM & 0.670 & 0.688 & -  \\
					Gupta, R. & graph & RVM & 0.670 & 0.690 & - \\
					Chao, H. et al. & MFM & CapsNet & 0.6828 & 0.6673 & 0.6725 \\
					\textbf{Ours} & - & ScalingNet & \textbf{0.6999} & \textbf{0.7113} & \textbf{0.7078} \\
					\bottomrule
				\end{tabularx}
			\end{table}
%			\end{sidewaystable}
			
			In Table \ref{tab:result}, Chao et al.\cite{2019Emotion} extract MFM features and use CapsNet as a classifier for emotion recognition. Chen et al.\cite{2019A} use H-AAT-BGRU to classify emotions. Li et al.\cite{2015EEG} and Yang et al.\cite{8683290} use SVM for classification by extracting DBN features and VAE features respectively. Gupta, R\cite{Gupta2016Relevance} use graph-theoretic features and RVM for classification.
			
			The results \ref{tab:result} show that the accuracy of the proposed method in this paper is 69.99\%, 71.13\%, and 70.78\% for arousal, valence, and dominance, respectively, which are both higher than the previous state-of-the-art studies. It indicates that the proposed ScalingNet architecture is effective and feasible for EEG  data based emotion recognition. Noticeably, its performance achieves the state-of-the-are result, but without any hand-interaction.
			
	\section{Discussion}
		%设计实验讨论参数、原理、现象，最后再引出咱们工作的优势
		% 讨论参数特点，反推背后工作机理、下结论支持咱们前面的设想 / 不要写太多，主要得支撑咱们的论点：东西好、有原理
		% todo 有什么可以讨论要进一步讨论
		In this section, we have elaborately designed a series of experiments to explore the properties of the scaling layer and ScalingNet, visualize the data-driven spectrogram-like feature maps extracted by scaling layers to explore its interpretability, and verify it's contribution through ablation experiments. 
	
		Since the scaling layer handles any length of EEG signals without requiring data alignment, we can arbitrarily adjust the length of the basic kernel $weight$ to explore the relationship between model capacity and its representational capacity.	We explore the relationship through observing the emotion recognition performance of ScalingNet with different setups of scaling layers. In the experiments, we deliberately select several representative parameters of the basic kernel $weight$ in scaling layers. The results is shown in Table \ref{tab:relationship}. 
		
		We can observe that the representational capacity attains the best at the model capacity of setting the length of $weight$ to 33. Obviously, the 33 is related to the DEAP dataset, and here are more interested in the Table \ref{tab:relationship} itself.
		
		\begin{table}[htbp]
			\centering
			\caption{The relationship of scaling layers between its model capacity and its representational capacity under the architecture of ScalingNet and the dataset of DEAP.}
			\label{tab:relationship}
			\begin{tabularx}{0.95\textwidth}{c | >{\centering\arraybackslash}X >{\centering\arraybackslash}X >{\centering\arraybackslash}X}
				\toprule
				\multirow{2}{*}{Length of $weight$} & \multicolumn{3}{c}{\textbf{Accuracy}} \\
				%\cline{2-4} 
				 & Arousal & Valance & Dominance \\
				\midrule
				129 & 0.6659 & 0.6778 & 0.6731 \\
				65 & 0.6773 & 0.6844 & 0.6886 \\
				63 & 0.6642 & 0.6882 & 0.6902 \\
				33 & \textbf{0.6999} & \textbf{0.7113} & \textbf{0.7078} \\
				17 & 0.6711 & 0.6726 & 0.6995 \\
				\bottomrule
			\end{tabularx}
		\end{table}
		
\begin{figure}[htbp]
			\centerline{
				\subfigure[lower frequency-like energy]{\includegraphics[width=0.45\textwidth]{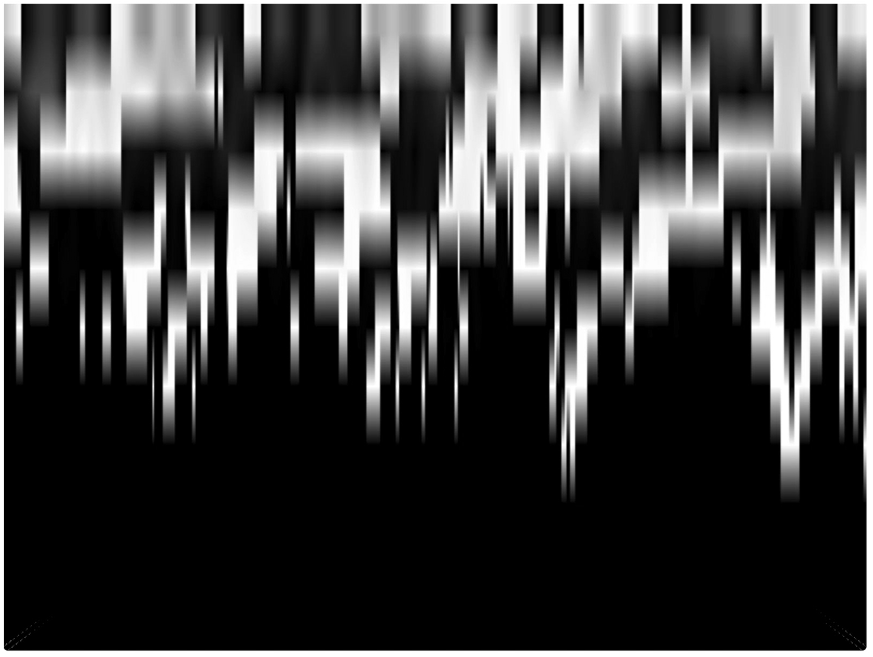}}
				\\
				\subfigure[higher frequency-like energy]{\includegraphics[width=0.45\textwidth]{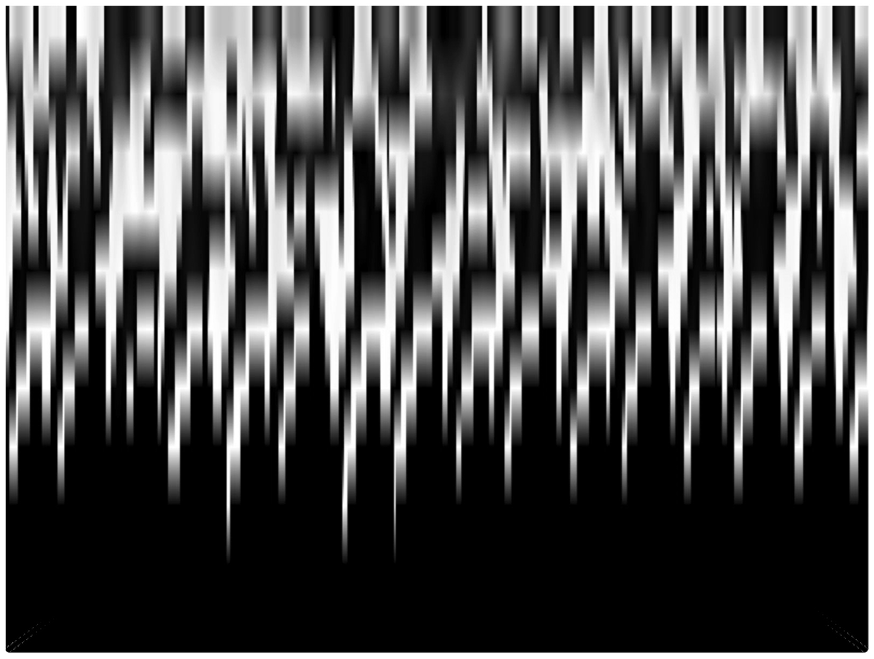}}
			}
			\caption{Data-driven spectrogram-like feature maps extracted by scaling layers under the architecture of ScalingNet and the dataset of DEAP.}
			\label{fig:featureMap}
		\end{figure}

		In addition, we visualize the data-driven spectrogram-like feature maps extracted by scaling layers under the architecture of ScalingNet and the dataset of DEAP. The visualized data-driven spectrogram-like feature maps are shown in Figure \ref{fig:featureMap}, where the horizontal axis denotes sampling points and the vertical axis denotes the frequency-like dimension, i.e. the time and scaling levels. We can observe that Figure \ref{fig:featureMap} (a) contains more low frequency-like energy and (b) contains more high frequency-like energy, it all starts with that one data-driven pattern that used to generate scaled kernels to extract useful information. These learned useful information contained in the data-driven spectrogram-like feature maps are aggregated by followed layers and used for downstream tasks.  
				
		In order to verify the contribution of the proposed scaling layer, ablation experiments are also considered. The results of ablation experiments are shown in Table \ref{tab:ablation}. In the ablation experiments, we compare the scaling layer with the convolutional layer to explore their feature extraction capability for EEG signals. We explore the capability through observing the emotion recognition performance with replacing the scaling layer of ScalingNet by the convolutional layer.We can observe that the scaling layers play an important role in ScalingNet. It also indicates that the scaling layer extract more robust feature for EEG signals with better generalization performance.
		
		\begin{table}[htbp]
			\centering
			\caption{The ablation experiments under the same backend architecture and the dataset of DEAP.}
			\label{tab:ablation}
			\begin{tabularx}{0.95\textwidth}{c | >{\centering\arraybackslash}X >{\centering\arraybackslash}X >{\centering\arraybackslash}X}
				\toprule
				\multirow{2}{*}{Feature extractor} & \multicolumn{3}{c}{\textbf{Accuracy}} \\ 
				%\cline{2-4} 
				 & Arousal & Valance & Dominance\\
				\midrule
				convolutional layer & 0.6574 & 0.6641 & 0.6628 \\
				scaling layer & \textbf{0.6999} & \textbf{0.7113} & \textbf{0.7078} \\
				\bottomrule
			\end{tabularx}
		\end{table}
				
	\section{Conclusion}
		%总结贡献、性能、方法间顾、未来
		% 结论胡景钊写
		% todo 做了什么、动机原理是什么、证明了前面讲的东西并依此解决了存在的问题、重申结果、【畅想未来】
		We have presented the scaling layer and ScalingNet, a novel convolutional layer for extracting a spectrogram-like feature map from raw signals and a neural network that operates on raw EEG data for classification, leveraging dynamically generated convolutional kernels by scaling from one data-driven pattern. We demonstrate that it can automatically adaptive extracting robust data-driven spectrogram-like feature maps and successfully applied to raw EEG data based emotion recognition. Thus it addresses many shortcomings of prior methods based on hand-extracted features with strong hypotheses or their approximations. Our ScalingNet models leveraging scaling layers have successfully achieved state-of-the-art performance across the well-established emotion recognition benchmarks.
		
	\section*{Acknowledgment}
		% 致谢资金支持，或需要感谢的人和物
		% todo 记得填充致谢
		This work was supported by the National Key Research and Development Program of China under grant 2017YFB1002504.

\bibliographystyle{unsrt}  
\bibliography{references}  %%% Remove comment to use the external .bib file (using bibtex).

\end{document}